\documentclass[12pt]{iopart}

\usepackage{iopams}
\usepackage{amssymb,amsfonts}
\usepackage{graphicx}
\usepackage[english]{babel}
\usepackage{mathrsfs}
\usepackage{rotate}

\begin{document}

\title[Chaotic and regular motion in a time-dependent barred galaxy
model]{Interplay Between Chaotic and Regular Motion in a Time-Dependent Barred Galaxy Model}

\author{T Manos$^{1,2,\dag}$, T Bountis$^{3,\ddag}$ and Ch Skokos$^{4,\S}$}

\address{$^1$ Center for Applied Mathematics and Theoretical Physics,
  University of Maribor, Krekova 2, SI-2000, Maribor, Slovenia}
\address{$^2$ School of Applied Sciences, University of Nova Gorica,
  Vipavska 11c, SI-5270, Ajdo\v{v}\v{s}cina, Slovenia}
\address{$^3$ Department of Mathematics and Center for Research and
  Applications of Nonlinear Systems (CRANS), University of Patras,
  GR-26500, Rion, Patras, Greece}
\address{$^4$ Section of Astrophysics, Astronomy and Mechanics,
  Physics Department, Aristotle University of Thessaloniki, GR-54124,
  Thessaloniki, Greece}
\ead{$\dag$thanos.manos@gmail.com, thanos.manos@uni-mb.si;
  $\ddag$bountis@math.upatras.gr;$\S$hskokos@auth.gr}

\begin{abstract}
  We study the distinction and quantification of chaotic and regular
  motion in a time-dependent Hamiltonian barred galaxy
  model. Recently, a strong correlation was found between the strength
  of the bar and the presence of chaotic motion in this system, as models with relatively  strong bars were shown to exhibit
  stronger chaotic behavior compared to those having a weaker bar
  component. Here, we attempt to further explore this connection by
  studying the interplay between chaotic and regular behavior of star
  orbits when the parameters of the model evolve in time. This happens
  for example when one introduces linear time dependence in the mass
  parameters of the model to mimic, in some general sense, the effect
  of self-consistent interactions of the actual N-body problem. We thus
  observe, in this simple time-dependent model also, that the increase
  of the bar's mass leads to an increase of the system's
  chaoticity. We propose a new way of using the Generalized Alignment
  Index (GALI) method as a reliable criterion to estimate
  the relative fraction of chaotic vs.~regular orbits in such time-dependent potentials, which proves to be much more efficient than the computation of Lyapunov exponents. In
  particular, GALI is able to capture subtle changes in the nature of an
  orbit (or ensemble of orbits) even for relatively small time
  intervals, which makes it ideal for detecting dynamical transitions in time-dependent
  systems.
\end{abstract}
\maketitle

\section{Introduction}
\label{Intro}

The study of chaotic and regular properties of the motion in Hamiltonian
systems constitutes a vast area of research in the field of nonlinear
dynamics. Since the early 1960's, several methods and tools for the fast and accurate detection of the nature of orbits have been proposed and applied to this end in a great number of publications. One may refer e.g.~to the pioneering paper by H\'{e}non and Heiles \cite{HenHei:1964}, where the Poincar\'{e} Surface of Section (PSS) \cite[section 1.2b]{LL} was used to reveal the chaotic properties of a non-integrable 2 degree of freedom (2 dof) Hamiltonian system. Of great importance in this direction was also the algorithm proposed by Benettin and co-workers \cite{Benettin1980a,Benettin1980b,SkoLE:2010} regarding the computation of the full spectrum of Lyapunov exponents (LEs) associated with the time evolution of deviation vectors from a reference orbit, which applies to dynamical systems of arbitrary dimension. More recently, other related methods have been proposed in the literature, like the ``Fast Lyapunov Indicator"
\cite{FroLeGo:1997,FroLe:1998} and the ``Mean Exponential Growth of
Nearby Orbits" (MEGNO) \cite{CinSimo:2000,CinGioSimo:2003}, while
there have also been approaches focusing on the time series
constructed by the coordinates of each orbit, like the ``Frequency Map
Analysis" \cite{Las:1990,LasFroCel:1992,Las:1993} and the ``0-1" test
\cite{GM04,GM09a,GM09b}. Interesting accounts of these methods can be
found in \cite{Con_spr}, as well as in a more recent review paper
\cite{SkoLE:2010}.

A novel, very efficient method based on the evolution of $k\geq 2$ initially linearly independent deviation vectors is provided by the so-called ``Generalized ALignment Indices" (GALI or GALI$_k$ spectrum) introduced in \cite{SBA:2007} as a generalization of the ``Smaller ALingment Index" (SALI)
\cite{Sk_sali:2001,SABV:2003a,SABV:2004}. The major advantage of the
GALI method is that it follows the evolution of two or more deviation
vectors and is thus able to extract more information about the
complexity of the motion, yielding i.e.~the dimensionality of the
invariant torus on which a regular orbit lies and predicting
faster the chaotic nature of trajectories \cite{SBA:2008,BouManChris:2010,ManRuf:2011}.
To date, the GALI and the SALI indices have been successfully applied to a wide variety of autonomous (i.e.~explicitly time-independent) conservative flows and maps (see e.g.~ \cite{PBS:2004,SESS,BouSko:2006,AntBou:2006,ABS:2006,CDLMV,VHC,KEV,MSAB:2008,Stra2009,BP09,Macek2010,SG10,MA11,BCSV12,MSA:2012,BCSPV12,GES12}). A concise review of the theory and applications of both the SALI and
GALI methods can be found in \cite[Chapter 5]{BS12}.

The motivation of the current work is twofold: First, we wish to
investigate the different dynamical properties of a non-autonomous
galactic potential, whose time-dependence could mimic certain
realistic general trends arising in barred N-body galaxy
simulations. Our second main goal is to explore the advantages of the
GALI method, over the more traditional LEs, in detecting dynamical transitions in Hamiltonian systems, whose equations of motion are explicitly time-dependent.

There are, of course, several studies of time-dependent (TD) galactic
and cosmological models in the literature, which use different tools
to identify the chaotic vs.~regular nature of orbits. Defining the
orbital complexity $n(k)$, of an orbital segment as the number of
frequencies in its discrete Fourier spectrum that contain a
$k$-fraction of its total power \cite{KandEckBra:1997}, one may
compare $n(k)$ with the short-time evolution of the LEs for TD models \cite{Sio_etal:1998}. In \cite{KandDru:98} the case of a cosmological model is discussed, where orbits may experience regular and/or chaotic motion during their time evolution, while in \cite{SioKand:2000} the effects of a black hole, friction, noise and periodic driving are studied on a triaxial elliptic galaxy model, in which a type of transient chaos was found caused by a damped,
oscillatory component \cite{KandVassSid:2003,TerzKand:2004}. Finally,
in \cite{Sid:2009} the so-called ``pattern method" was used to study a
H\'{e}non-Heiles potential to which an exponential function of time is
added, while the dynamics of some simple TD galactic models was
investigated in \cite{CarPap:2003,Zot:2012}.


We recall here that in conservative systems the asymptotic nature of
an orbit may be either periodic, quasiperiodic or chaotic. In the
latter case, however, it may take a very long time before one can
safely claim that a ``final" state is reached, depending on the local
dynamical properties, which may be characterized by ``strong or weak
chaos". In \cite{Kats:2011a,Kats:2011b,MSA:2012} the dynamics in the vicinity of periodic orbits in conservative systems was studied by means of the maximum Lyapunov exponent (MLE) and the GALI. Here we explore the advantages of the GALI method and compare its predictions with what one finds using more traditional methods like the computation of the MLE also for TD systems.

In particular, we focus our attention on the dynamics of a barred
galaxy model containing a disc and a bulge component, which is a widely
accepted model for real barred galaxies. In the spirit of a mean field approach, we consider the motion of stars (represented by point particles) in this potential. The richness of the dynamics of the time-independent (TI) version of this model has been extensively studied in terms of: (a) the detection of periodic orbits and the analysis of their stability (see e.g.~\cite{ABMP,Pfe:1984a,SPA02a,SPA02b,PSA02,PSA03a,PSA03b,SPPV05}), (b) the estimation of the relative fraction of chaotic vs.~regular orbits
\cite{MSAB:2008,MA:2009,MA11}, and (c) the statistical distributions
of orbital coordinates described by $q$-Gaussian distribution functions
\cite{BouManAnt:2012}.

Here, we extend the analysis by considering a TD version of this
model. More specifically, we allow some mass parameters of the
potential to vary linearly as functions of time. As expected, whether
we study a 2 dof or 3 dof version of the model, these variations can
change the stability properties of periodic orbits, ``dissolve'' islands
of regular motion and alter the structure of phase space in very complicated
ways. Furthermore, in the TD case, the vast majority of dynamical transitions of phase space orbits cannot be claimed to be due to stickiness phenomena or ordinary diffusion to different regimes, as expected for TI Hamiltonian systems.

Recently, it was found in the TI case that the relative fraction
of chaotic orbits grows as the bar's strength increases \cite{MA11}. A
question therefore arises, whether a similar correlation holds in the
TD model, in the presence of ``realistic" trends, which permit the
mean field potential to vary in a way that is compatible with
self-consistent N-body simulations regarding several components of the
system.

Clearly, the analysis of the full N-body problem describes much better
the galactic evolution and captures in great detail the different
stellar structures present in the dynamics. However, there are serious
difficulties and limitations when one tries to apply dynamical chaos
detectors to such ``realistic" many-particle systems due to the lack
of sufficient orbital information during the time evolution. For this
reason, many researchers prefer to use mean field potentials that are
``frozen" in time and study the properties at specific snapshots of
the simulations \cite{Kalapo:2008,HaKa:2009}.

Keeping in mind that a barred galaxy experiences several dynamical
transitions in different epochs that cannot be easily incorporated in
our TD mean field potential, we shall proceed by making some helpful
assumptions in an attempt to understand the behavior of such widely used chaos
detectors as the GALI and the MLE. Thus, we will treat here two
very general dynamical trends known to occur in barred galaxies: In
the first scenario the mass of the bar component grows linearly in
time (at the expense of the disc mass). This increase may be caused by
an exchange of angular momentum with the disc (outer parts gain
momentum from the inner parts), as has already been observed in N-body
simulations (see e.g.~\cite{AthMiss:2002,Ath:2003}). The fundamental
trend in this case is that bars generally grow stronger in time,
become more elongated and massive and eventually slow down.  We will also
consider the inverse scenario, where the bar gets weaker making the disc
more massive as time evolves (see e.g.~\cite{Combes:2008,Combes:2011}).

The paper is organized as follows: In section \ref{Model_gal} we
present the TD barred galaxy model used in our study, while section
\ref{methods} is devoted to the description of the numerical
methods employed for the computation of the MLE and the
GALIs. Section \ref{results} contains the main numerical results of
the paper. A detailed investigation of the dynamics of particular
orbits in a 2 dof version of our model is performed in section
\ref{2DOF}, while orbits of the full, 3 dof model, are considered in
section \ref{sec:3dof}. A global investigation of the dynamics of our
TD galactic model is given in section \ref{sec:global}. Finally, in
section \ref{sec:conclusions} the main conclusions of our work are
presented.

\section{The model potential}
\label{Model_gal}

Let us consider the following TD 3 dof Hamiltonian function which
determines the motion of a star in a 3 dimensional rotating barred
galaxy:
\begin{equation}\label{eq:Hamilton}
  H=\frac{1}{2} (p_{x}^{2}+p_{y}^{2}+p_{z}^{2})+ V(x,y,z,t) -
  \Omega_{b} (xp_{y}-yp_{x}).
\end{equation}
The bar rotates around its $z$--axis (short axis), while the $x$
direction is along the major axis and the $y$ along the intermediate
axis of the bar. The $p_{x}$, $p_{y}$ and $p_{z}$ are the canonically
conjugate momenta, $V$ is the potential, $\Omega_{b}$ represents the
pattern speed of the bar and $H$ is the total energy of the orbit in
the rotating frame of reference (equal to the Jacobi constant in the
TI case).

The corresponding equations of motion are:
\begin{equation}\label{eq_motion}
\begin{array}{lcl}
  \dot{x} &=& \displaystyle p_{x} + \Omega_{b} y, \\
  \dot{y} &=& \displaystyle p_{y} - \Omega_{b} x,  \\
  \dot{z} &=& \displaystyle p_{z}, \\
  \dot{p_{x}} &=& \displaystyle -\frac{\partial V}{\partial x} + \Omega_{b} p_{y}, \\
  \dot{p_{y}} &=& \displaystyle -\frac{\partial V}{\partial y} - \Omega_{b} p_{x}, \\
  \dot{p_{z}} & =& \displaystyle -\frac{\partial V}{\partial z}, \\
\end{array}
\end{equation}
while the equations governing the evolution of a deviation vector
$\mathbf{w}=(\delta x,\delta y,\delta z,\delta p_{x},\delta
p_{y},\delta p_{z})$ needed for the calculation of the MLE
and the GALIs, are given by the variational equations:
\begin{equation}\label{eq_dev_vect}
\begin{array}{lcl}
  \dot{\delta x} &=& \displaystyle \delta p_{x} + \Omega_b \delta y,  \\
  \dot{\delta y} &=& \displaystyle \delta p_{y} + \Omega_b \delta x,  \\
  \dot{\delta z} &=& \displaystyle\delta p_{z},\\
  \dot{\delta p_{x}} &=& \displaystyle- \frac{\partial^2 V}{\partial x \partial x}\delta x -
  \frac{\partial^2 V}{\partial x \partial y}\delta
  y - \frac{\partial^2 V}{\partial x \partial z} \delta z + \Omega_{b} \delta p_{y}, \\
  \dot{\delta p_{y}} &=& \displaystyle- \frac{\partial^2 V}{\partial y \partial x}\delta x -
  \frac{\partial^2 V}{\partial y \partial y}\delta
  y - \frac{\partial^2 V}{\partial y \partial z} \delta z - \Omega_{b} \delta p_{x}, \\
  \dot{\delta p_{z}} &=& \displaystyle- \frac{\partial^2 V}{\partial z \partial x}\delta x -
  \frac{\partial^2 V}{\partial z \partial y}\delta y - \frac{\partial^2 V}{\partial z \partial z} \delta z. \\
\end{array}
\end{equation}

The potential $V$ of our model consists of three components:
\begin{enumerate}
\item[(a)] A triaxial Ferrers bar \cite{Fer}, the density
  $\rho(x,y,z)$ of which is given by:
\begin{eqnarray}
  \rho(x,y,z) = \left\{
  \begin{array}{l l}
    \rho_{c}(1-m^{2})^{2}& \quad \textrm{if} \quad m<1,\\
    \quad 0& \quad \textrm{if} \quad m \geq 1,\\
  \end{array} \right.
\end{eqnarray}
where $\rho_{c}=\frac{105}{32\pi}\frac{G M_{B}(t)}{abc}$ is the
central density, $M_{B}(t)$ is the mass of the bar which changes in
time, and
$m^{2}=\frac{x^{2}}{a^{2}}+\frac{y^{2}}{b^{2}}+\frac{z^{2}}{c^{2}}$,
$a>b>c> 0$, with $a,b$ and $c$ being the semi-axes of the ellipsoidal bar. The corresponding
potential is:
\begin{equation}\label{Ferr_pot}
    V_{B}= -\pi Gabc \frac{\rho_{c}}{3}\int_{\lambda}^{\infty}
    \frac{du}{\Delta (u)} (1-m^{2}(u))^{3},
\end{equation}
where $G$ is the gravitational constant (set equal to unity here),
$m^{2}(u)=\frac{x^{2}}{a^{2}+u}+\frac{y^{2}}{b^{2}+u}+\frac{z^{2}}{c^{2}+u}$,
$\Delta^{2} (u)=({a^{2}+u})({b^{2}+u})({c^{2}+u})$, and $\lambda$ is
the unique positive solution of $m^{2}(\lambda)=1$, outside of the bar
($m \geq 1$), while $\lambda=0$ inside the bar. The analytical
expression of the corresponding forces are given in \cite{Pfe:1984a}.

\item[(b)] A bulge, modeled by a Plummer sphere \cite{Plum} whose
  potential is:
\begin{equation}\label{Plum_sphere}
  V_S=-\frac{G M_{S}}{\sqrt{x^{2}+y^{2}+z^{2}+\epsilon_{s}^{2}}},
\end{equation}
where $\epsilon_{s}$ is the scale-length of the bulge and $M_{S}$ is
its (constant) mass.
\item[(c)] A disc, represented by the Miyamoto-Nagai potential
  \cite{Miy.1975}:
 \begin{equation}\label{Miy_disc}
   V_D=- \frac{GM_{D}(t)}{\sqrt{x^{2}+y^{2}+(A+\sqrt{z^{2}+B^{2}})^{2}}},
\end{equation}
where $A$ and $B$ are its horizontal and vertical scale-lengths, and
the mass of the disc $M_{D}(t)$ changes in time so that the total mass
of the system is kept constant.
\end{enumerate}

The model's parameters have the following constant values: $G=1$,
$\Omega_{b}$=0.054 (54 $km\cdot sec^{-1} \cdot kpc^{-1}$), $a$=6,
$b=$1.5, $c$=0.6, $A$=3, $B$=1, $M_{S}$=0.08, while the initial values
of the bar and disc masses are $M_{B}(0)$=0.1 and $M_{D}(0)$=0.82,
respectively. The units used are: 1 $kpc$ (length), 1000 $km\cdot
sec^{-1}$ (velocity), 1 $Myr$ (time), $2 \times 10^{11} M_{\bigodot}$
(mass). The total mass $M_{S}+M_{D}(t)+M_{B}(t)$ is set equal to 1 and
since the bulge's mass $M_S$ is kept constant, the disc's mass
$M_D(t)$ is varied as $M_D(t)=1-(M_S+M_B(t))$. The rate of the mass
variation of the bar is chosen to be linear according to the law:
\begin{equation}\label{MBlinlaw}
  M_B(t)=M_B(t_0=0)+\alpha t,
\end{equation}
where the proportionality constant is $\alpha>0$ or $\alpha<0$
respectively, if the mass of the bar increases or decreases in time.

In order to measure the time variation of the bar's strength we
calculate the quantity \cite{ButaBK03, ButaLS04}:
\begin{equation}\label{Qt}
  Q_t(r) =  \displaystyle \left(\frac{\partial \Phi(r,\theta)}{\partial \theta}\right)_{\mbox{max}}\cdot\displaystyle \left(r \frac{\partial \Phi_0}{\partial r}\right)^{-1},
\end{equation}
which estimates the relative strength of the non-axisymmetric forces.
In the above expression, $\Phi$ is the potential on the symmetry plane
$z=0$ expressed in polar coordinates $(r,\theta)$, $\Phi_0$ is its
axisymmetric part, while the maximum in the first term on the right hand side of (\ref{Qt}) is calculated over all values of the azimuthal angle $\theta$. The maximum value of $Q_t(r)$ over all radii shorter than the bar extent,
termed $Q_b$, can be used as a measure of the bar's strength.

It is clear, of course, that the variation of the bar's strength modifies
the values of several parameters and yields richer information about
the dynamics of a self-consistent model. N-body simulations show that
in general, variations of the bar's mass also change the mass ratios of
the model's components, the bar's shape and the pattern speed of the galaxy.
Hence, if one wishes to use a mean field potential to
``mimic'' a self-consistent model as accurately as possible, one should
allow for all the parameters that describe the bar (together with all
other axisymmetric components) to depend on time, assuming that
the laws of such dependence were explicitly known. In our case,
however, we adopt a simpler approach and vary only the masses of
the bar and the disc, as a first step towards investigating such models when
time dependent parameters are taken into account. Thus, we do not pretend to be able
to reproduce the exact dynamical evolution of a realistic galactic simulation.
Rather, we wish to understand the effects of time dependence on the
general features of barred galaxy models and compare the efficiency
of indicators like the GALIs and the MLE in helping us unravel the
secrets of the dynamics in such problems.

\section{Computational methods}\label{methods}

In order to estimate the value of the MLE, $\lambda_1$, of a
particular orbit we follow the evolution of the orbit and
a deviation vector $\mathbf{w}$ from it, by numerically solving the
set of equations (\ref{eq_motion}) and (\ref{eq_dev_vect})
respectively. For this task we use a Runge-Kutta method of order 4
with a sufficiently small time step (typically of the order of $\tau
\approx 10^{-2}$), which guarantees the accuracy of our computations
(i.e.~the use of the half time step does not practically change our
results).

The ordinary differential equations (ODEs) (\ref{eq_motion}) can be
solved independently from equations (\ref{eq_dev_vect}). On the other
hand, the latter set of equations, governing the evolution of a
deviation vector, has to be solved simultaneously with equations
(\ref{eq_motion}), because the second derivatives of the potential
$V$, appearing in the right hand side of (\ref{eq_dev_vect}), depend
explicitly on the solution of (\ref{eq_motion}). Note that
(\ref{eq_motion}) constitutes a non-autonomous set of ODEs because the
derivatives of $V$ depend explicitly on time. Although one could
transform (\ref{eq_motion}) (and consequently (\ref{eq_dev_vect})) to
an equivalent autonomous system of ODEs by considering time $t$ as an
additional coordinate (see e.g.~\cite[section 1.2b]{LL}), this
approach is not particularly helpful, and is better to be avoided
\cite{GrySzla:1995}.

So, in order to compute the MLE and the GALIs we numerically solve the
time-dependent set of ODEs (\ref{eq_motion}) and (\ref{eq_dev_vect}). Then, according to \cite{BGS76,CGG78,Benettin1980a,Benettin1980b} the MLE $\lambda_1$ is defined as:
\begin{equation}
\label{LE}
\lambda_1 =\lim_{t \rightarrow \infty} \sigma_1(t),
\end{equation}
where:
\begin{equation}
\label{sigma_1}
\sigma_1(t)= \frac{1}{t}
\ln \frac{\|\mathbf{w}(t)\|}{\|\mathbf{w}(0)\|},
\end{equation}
is the so-called ``finite time MLE'', with $\|\mathbf{w}(0)\|$ and
$\|\mathbf{w}(t)\|$ being the Euclidean norm of the deviation vector
at times $t=0$ and $t>0$ respectively. A detailed description of the
numerical algorithm used for the evaluation of the MLE can be found in
\cite{SkoLE:2010}.

This computation can be used to distinguish between regular and chaotic orbits, since $\sigma_1(t)$ tends to zero (following a power law $\propto t^{-1}$) in the former case, and converges to a positive value in the latter. But Hamiltonian (\ref{eq:Hamilton}) is TD, which means that its orbits could change their dynamical behavior from regular to chaotic and vice versa, over different time intervals of their evolution. In such cases, the computation of the MLE (\ref{LE}) might not be able to identify the various dynamical phases of the
orbits, since by definition it characterizes the asymptotic behavior of an orbit.

In order to avoid such problems in our study, we also turn to the use of the GALI method of chaos detection \cite{SBA:2007}. The GALI index of order $k$
(GALI$_k$) is determined through the evolution of $2 \leq k \leq N$
initially linearly independent deviation vectors $\textbf{w}_i(0)$, $i
= 1,2,\ldots,k$, with $N$ denoting the dimensionality of the phase space of
our system. Thus, apart from solving (\ref{eq_motion}), which determines the evolution of an orbit, we have to simultaneously solve (\ref{eq_dev_vect}) for each one of the $k$ deviation vectors. Then, according to \cite{SBA:2007}, GALI$_k$ is defined as the volume of the $k$-parallelogram having as edges the $k$ unit deviation vectors
$\hat{\textbf{w}}_i(t)=\textbf{w}_i(t)/\|\textbf{w}_i(t)\|$, $i =
1,2,...,k$. It can be shown, that this volume is equal to the norm of
the wedge product (denoted by $\wedge$) of these vectors:
\begin{equation}\label{GALI:0}
  \textrm{GALI}_{k}(t)=\parallel \hat{\textbf{w}}_{1}(t) \wedge \hat{\textbf{w}}_{2}(t) \wedge \ldots \wedge \hat{\textbf{w}}_{p}(t) \parallel.
\end{equation}
We note that in the above equation the $k$ deviation vectors are
normalized but their directions are kept intact. In practice, we apply
a numerical method for calculating this norm, which is based on the
singular value decomposition of an appropriate matrix
\cite{AntBouLDI:2006,SBA:2008}.

The behavior of GALI$_k$ for regular and chaotic orbits was
theoretically studied in \cite{SBA:2007,SBA:2008}, where it was shown
that all GALI$_k(t)$ tend exponentially to zero for chaotic orbits, with
exponents that depend on the first $k$ LEs of the orbit. However, while this relation has been verified for TI systems, in the TD case studied here, the way the GALI exponential rates depend on the LEs is  less clear and certainly requires further investigation. In the case of regular orbits, GALI$_k$ remains practically constant and positive if $k$ is smaller or equal to the dimensionality of the torus on which the motion occurs, otherwise, it decreases to zero following a power law decay. We may, therefore, say that the GALIs do contain important geometric information about the tangent space of the orbits, in the sense that they identify the number of linearly independent deviation vectors in phase space. This information is especially useful near quasiperiodic orbits, where it helps us accurately determine the dimensionality of the associated torus \cite{SBA:2008}.

In order to use GALI$_k$ to capture the dynamical changes of orbits
in TD systems we apply the following procedure: Whenever GALI$_k$
reaches very small values (i.e.~GALI$_k \leq 10^{-8}$) we reinitialize
its computation by taking again $k$ new random orthonormal deviation
vectors, which means that we set again GALI$_k=1$. Then we let these
vectors evolve under the current dynamics. An exponential decrease of GALI$_k$
indicates chaotic behavior. Thus, the time $t_d$ needed for GALI$_k$
to become less than $10^{-8}$ can be used to identify epochs where the
orbit is chaotic or regular.

\section{Numerical results}\label{results}

\subsection{A typical  orbit of the 2 dof model}\label{2DOF}

Let us first start with the simple example of an orbit whose motion is
restricted in the 2 dimensional (2D) $(x,y)$ space (or 4D phase
space), of a 2 dof reduced version of the full 3 dof model, where
$z,p_z$ are set equal to zero at $t=0$, and remain zero at all
times. Furthermore, we shall assume that the bar component gets
stronger in time, according to (\ref{MBlinlaw}), in agreement with the
general trend in the dynamical evolution of barred galaxies
(e.g.~\cite{AthMiss:2002,Ath:2003}). In general, the fraction of the
chaotic component here is expected to increase as the bar gains mass
and becomes stronger \cite{MA11}. However, this does not necessarily
imply that the nature of all orbits remains unchanged in time; regular
orbits can become chaotic as time goes on, and vice versa. For this
system one can construct, at specific times, 2D phase portraits (similar to the PSS of the TI case) to visualize the dynamics and follow the changes
that orbits undergo.

Assuming that all other components except $M_B(t)$ and $M_D(t)$ remain
constant, we vary $M_B$ from $M_B(t_0=0)=0.1$ to $M_B(t_{final}=20000)=0.2$, yielding a proportionality constant $\alpha=5 \times 10^{-6}$ in (\ref{MBlinlaw}). The initial energy value for these parameters is set to be
$H_A(t_0)=E_A(t_0)=-0.2570$. The strength of the bar, measured by the
$Q_b$ parameter is initially $Q_b(t_0)=0.425$ and becomes
$Q_b(t_{final})= 0.6732$.

As a test case, we pick a specific orbit, which we call orbit A,
with initial condition $(x,y,p_x,p_y)=(0.0,1.0,0.16531,0.0)$. This
orbit is initially located inside an island of stability in the
system's phase space and is thus expected to remain regular, at
least for some time to come.

\begin{figure}
\centering
\includegraphics[width=4.9cm]{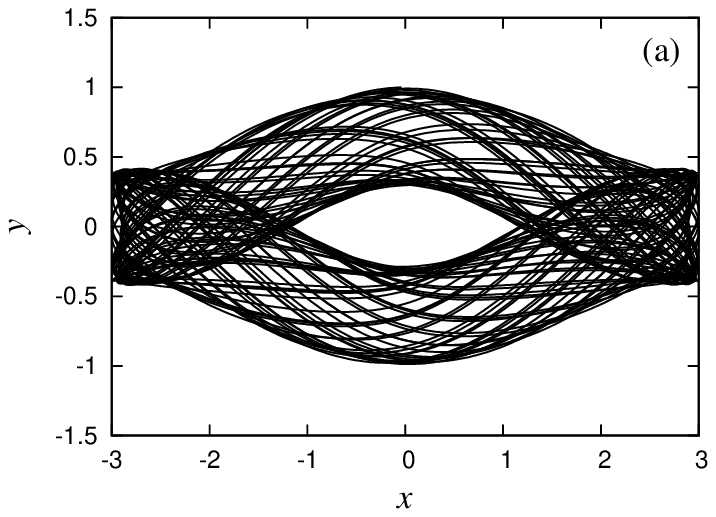}
\includegraphics[width=4.9cm]{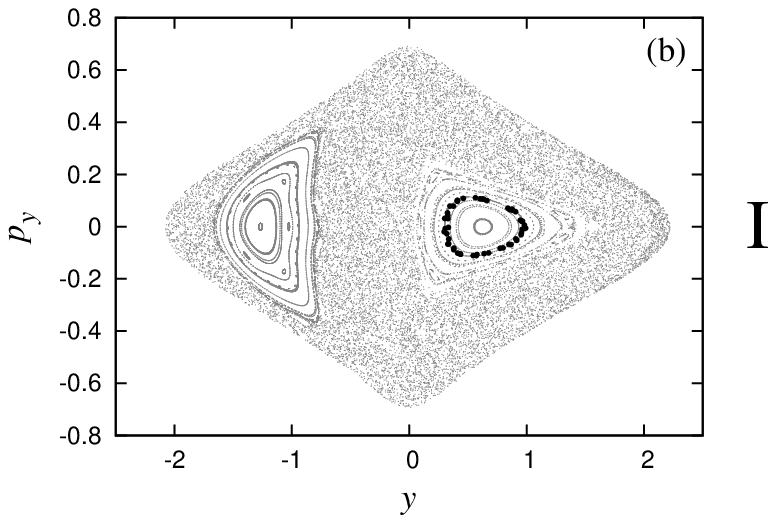}\\
\includegraphics[width=4.9cm]{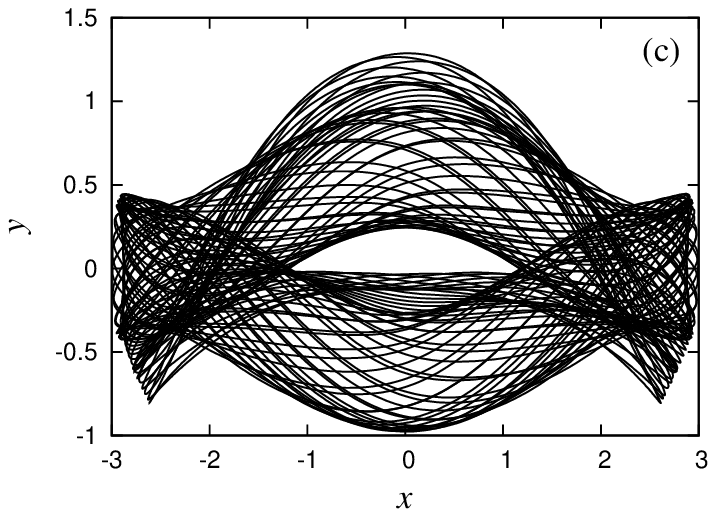}
\includegraphics[width=4.9cm]{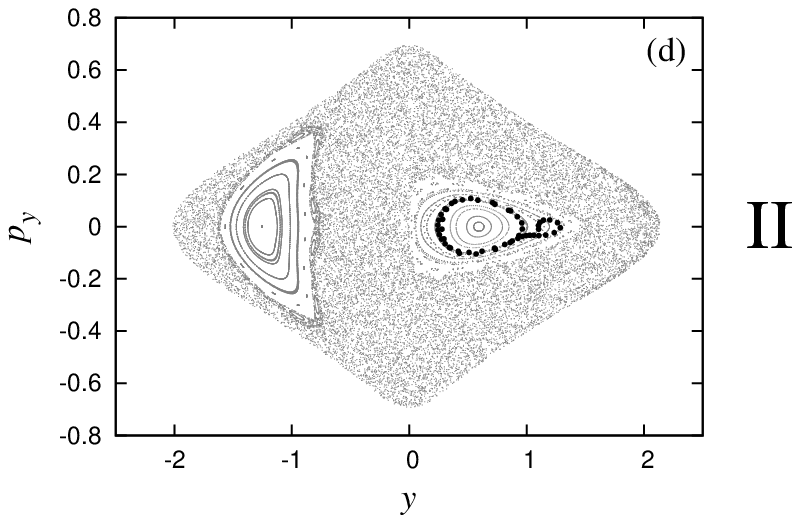}\\
\includegraphics[width=4.9cm]{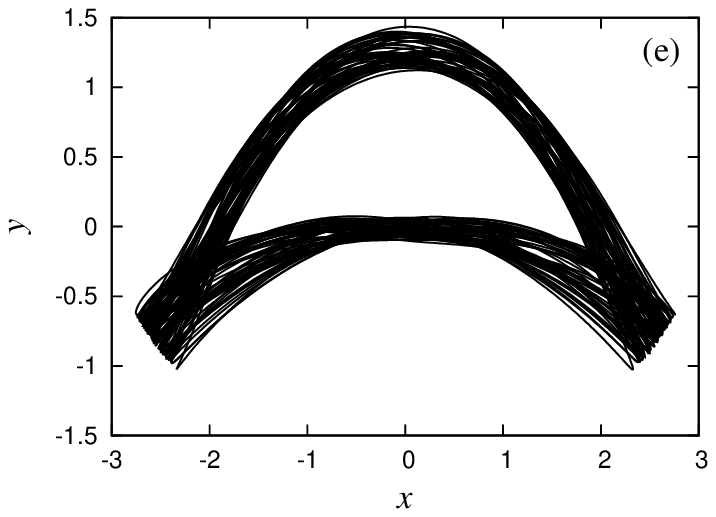}
\includegraphics[width=4.9cm]{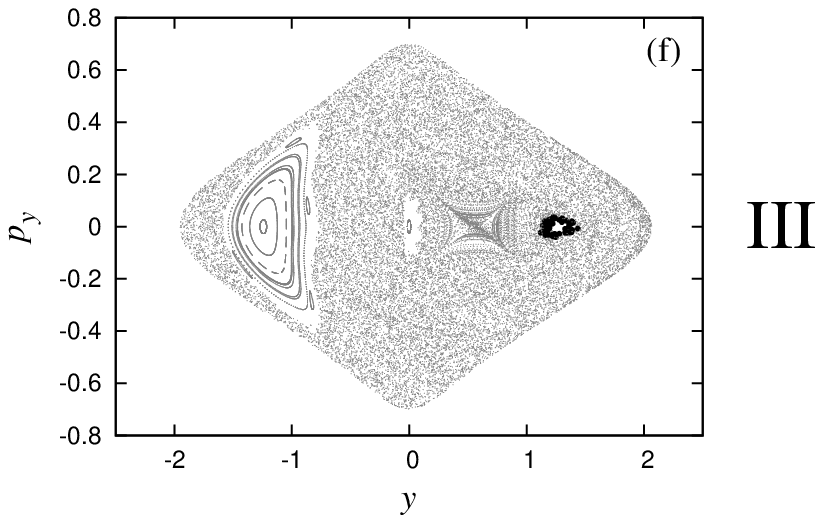}\\
\includegraphics[width=4.9cm]{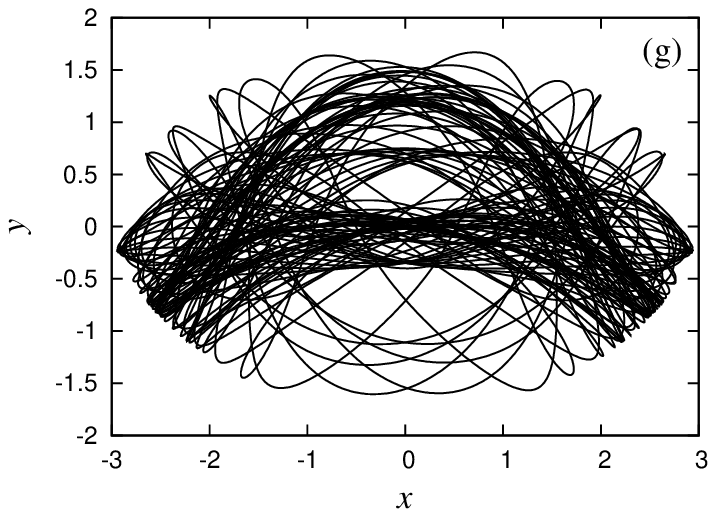}
\includegraphics[width=4.9cm]{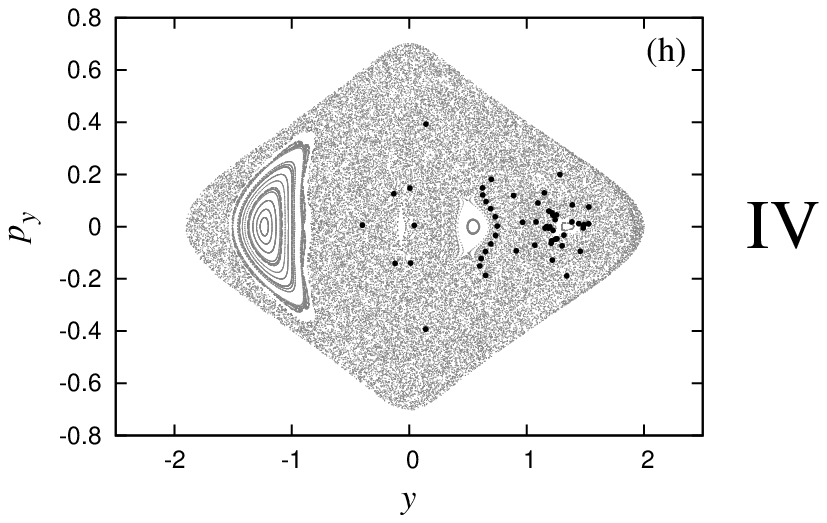}\\
\includegraphics[width=4.9cm]{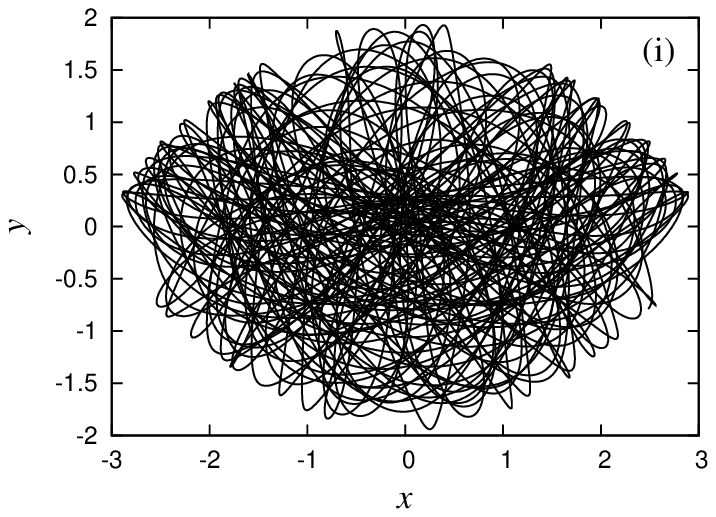}
\includegraphics[width=4.9cm]{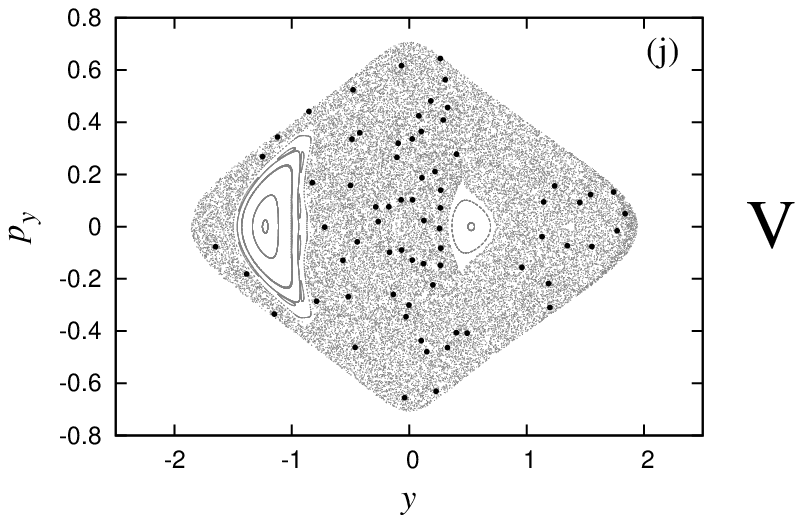}\\
\caption{Left column: projections of orbit A on the $(x,y)$
  configuration space for successive time intervals with 2500 time
  units length: $0\leq t \le 2500$ (first row, interval I), $2500< t
  \le 5000$ (second row, interval II), $5000< t \le 7500$ (third
  row, interval III), $7500< t \le 10000$ (fourth row, interval IV)
  and $10000< t \le 12500$ (fifth row, interval V). Note that $y$
  axis has not the same size in all panels. Right column: intersection
  points of orbit A with the PSS $x=0$, $p_x \geq 0$ for the same time
  intervals (black points). In order to get a clear picture of the
  structural evolution of the phase space, in each panel the PSS
  corresponding to the central time of each time interval ($t=1250$,
  $3750$, $6250$, $8750$ and $11250$, from top to bottom) is
  plotted in gray. The orbit is initially regular and drifts from one
  island of stability to another, until finally its dynamical nature
  can be characterized as chaotic.}
\label{2DOF_ORB_PSSs}
\end{figure}

In the left column figure \ref{2DOF_ORB_PSSs}
we show the projections of orbit A on the $(x,y)$ plane, while on the
right column we  plot its intersection points with a PSS defined by $x=0$,
$p_x \geq 0$ (black points) for five successive time
intervals, each having a duration of 2500 time units. In every panel
of the right column we plot in gray the PSS which corresponds to the
central time of each time interval, i.e.~$t=1250$, $t=3750$,
$t=6250$, $t=8750$ and $t=11250$, respectively from top to bottom.
Orbit A is initially located inside the right island of stability of
figure \ref{2DOF_ORB_PSSs}(b) and oscillates symmetrically around the
bar's major-axis, as shown in figure \ref{2DOF_ORB_PSSs}(a) for $0\leq
t \le 2500$ (interval I).  Then in figure \ref{2DOF_ORB_PSSs}(d) we
observe a first drift from the original island of stability to a
nearby one for $2500< t \le 5000$ (interval II). The morphology of
the orbit (figure \ref{2DOF_ORB_PSSs}(c)) changes at the same time to
a different shape. In figures \ref{2DOF_ORB_PSSs}(e),(f) this transition has fully taken place and now the motion occurs on an island different from the one it started on, but its regular nature is still preserved for $5000< t \le 7500$ (interval III). In the next time interval, however, $7500< t \le 10000$ (interval IV), we see a radical change in the orbit's morphology, which indicates the transition from regularity to chaoticity (figures \ref{2DOF_ORB_PSSs}(g),(h)). From figures \ref{2DOF_ORB_PSSs}(i),(j) we deduce that for $10000< t \le 12500$ (interval V), orbit A moves to regions away from the bar in the configuration space, enters the big chaotic sea on the PSS, and from then on shows no regular behavior, as it remains in this chaotic region for the rest of the integration time. We stress here again that the above dynamical transitions are not related to stickiness or ordinary diffusion phenomena as one finds in TI Hamiltonian systems.

\begin{figure}
\centering
\includegraphics[width=7.0cm]{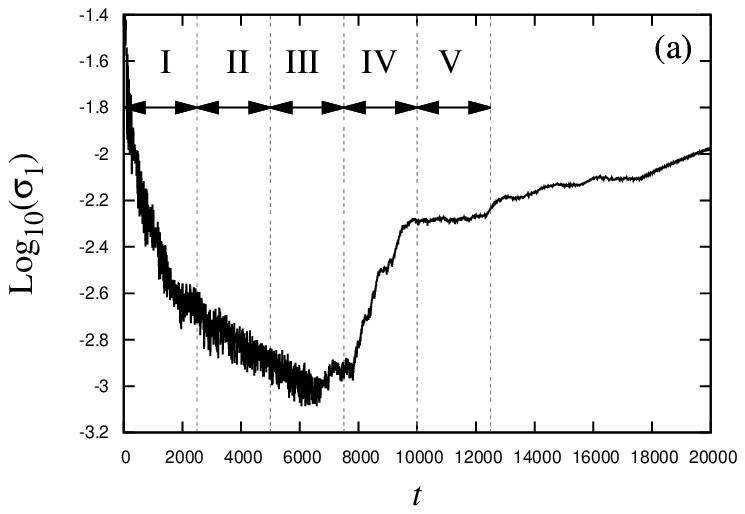}
\includegraphics[width=7.0cm]{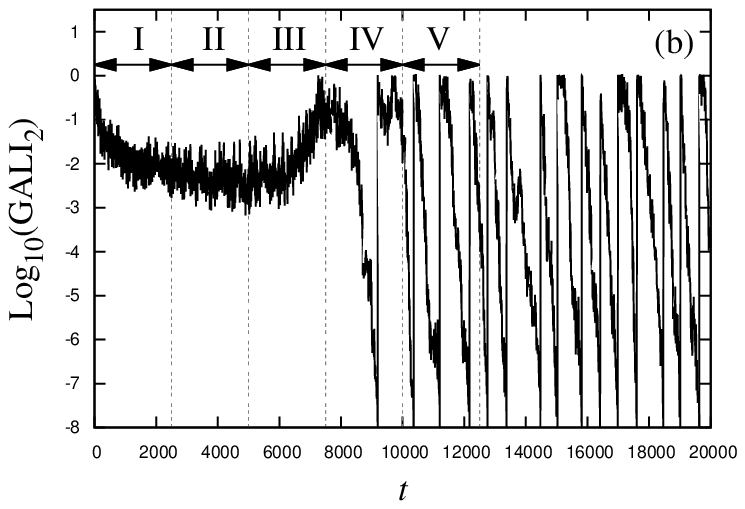}
\caption{Time evolution of the logarithm of (a) the finite time MLE
  $\sigma_1$, and (b) the reinitialized GALI$_2$ of orbit A. The five
  different time intervals I, II, III, IV and V, that correspond to
  the rows of figure \ref{2DOF_ORB_PSSs}, are located between the
  vertical dashed gray lines.}
\label{2DOF_MLE_GALI}
\end{figure}

In figure \ref{2DOF_MLE_GALI}(a)
we depict the time evolution of the finite time MLE $\sigma_1(t)$
(\ref{sigma_1}) of orbit A.  During the first parts of the motion
(intervals I and II), we clearly see a decay of $\sigma_1$ to zero
indicating the regular nature of the orbit, in accordance with the
results of figure \ref{2DOF_ORB_PSSs}. Later on, at the end of
interval III and mainly during interval IV, we witness a transient behavior
where $\sigma_1$ stops decaying and chaos arises. Then, in
interval V $\sigma_1$ remains positive and shows a tendency to slightly
increase, which clearly suggests that orbit A becomes more chaotic
as the bar's mass increases.

\begin{figure}
\centering
\vspace{1cm}
\includegraphics[width=5.0cm]{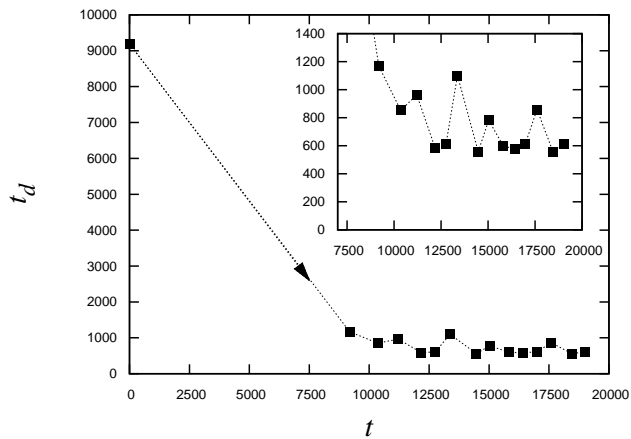}
\vspace{2cm}
\caption{The time intervals $t_d$ needed for the reinitialized GALI$_2$ to
  decrease from GALI$_2=1$ to GALI$_2\leq 10^{-8}$, as a function of
  the integration time $t$ of orbit A. The fluctuations of the $t_d$ values (see insert) reflect the inhomogeneity of the dynamics in the corresponding time intervals.}
\label{2DOF_GALI2zeroT}
\end{figure}

Let us now examine the behavior of GALI$_2$ for the same orbit. From
figure \ref{2DOF_MLE_GALI}(b) we see that GALI$_2$ oscillates around a
positive value during the time intervals I and II for which the orbit
is regular and decays exponentially to zero, becoming $\leq 10^{-8}$,
as soon as the orbit enters interval IV. In order to monitor the
dynamical changes of the orbit we reinitialize the
computation of GALI$_2$ as soon as GALI$_2 \leq 10^{-8}$, and
plot in figure \ref{2DOF_GALI2zeroT} the time $t_d$ needed for GALI$_2$ to decrease from GALI$_2=1$ to values smaller than $10^{-8}$ along the orbit's
evolution. This figure demonstrates that orbit A is
initially regular and its GALI$_2$ becomes $\leq 10^{-8}$ for the
first time after $t\approx 9500$ within interval IV. From that
point on the orbit remains chaotic as its reinitialized GALI$_2$
repeatedly falls to zero very fast, resulting in small $t_d$ values
($t_d \lesssim 1000$). This phase corresponds to the times that the
orbit wanders in the big chaotic sea of the PSS (see figures \ref{2DOF_ORB_PSSs}(h) and (j)).

The important observation here is that the initial transition of
orbit A from regularity to chaoticity and the subsequent variations of its dynamics are not distinctly captured by the evolution of the finite time MLE. Note from (\ref{LE}) and (\ref{sigma_1}) that the MLE represents a time-averaged quantity over the whole evolution of the orbit, and consequently cannot reveal detailed changes of the orbit's motion as it exits an island and wanders within a large chaotic sea. On the other hand, GALI$_2$ {\it does} reveal such changes as it depends only on the current state of the dynamics and not on the previous history of the orbit. In fact, these advantages of the GALIs in capturing even brief dynamical transitions become more evident in the following sections, where we study orbits in 3 dof TD models.

\subsection{Orbits of the 3 dof model}\label{sec:3dof}

After describing how successful the finite time MLE and the GALI are
in detecting changes in the chaotic vs.~regular nature of orbits in the
2 dof restriction of Hamiltonian (\ref{eq:Hamilton}), let us study
some representative cases of the full 3 dof problem.

\subsubsection{A case where the bar gets weaker in time}\label{sec:mass_decrease}

We now suppose that the bar's mass decreases linearly
in time from an initial value $M_B(t_0=0)=0.1$ to
$M_B(t_{final}=20000)=0.0$, following the law (\ref{MBlinlaw}) with
$\alpha=-5 \times 10^{-6}$ and study an orbit with
quite interesting behavior, which we call orbit B. Its initial
condition is $(x,y,z,p_x,p_y,p_z)=(0.3124,0.0,0.25,0.0,0.0)$, and its
initial energy $H_B(t_0)=E_B(t_0)=-0.429$.  In this case, the bar's
strength starts at $Q_b(t_0)=0.425$ and reaches the value
$Q_b(t_{final})= 0$ when there is no mass left at the bar component of
the model.

\begin{figure}
\centering
\includegraphics[width=5.0cm]{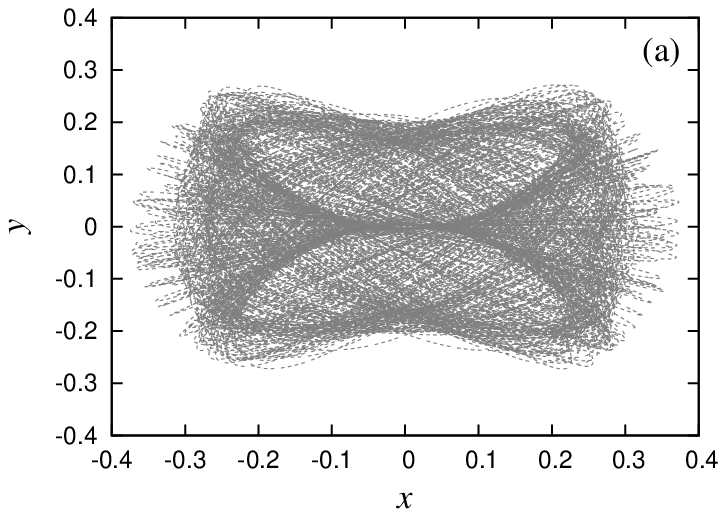}
\includegraphics[width=5.0cm]{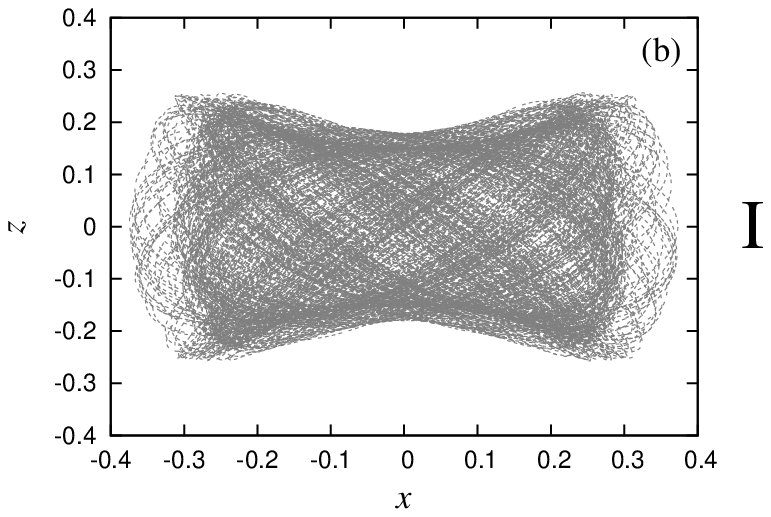}\\
\includegraphics[width=5.0cm]{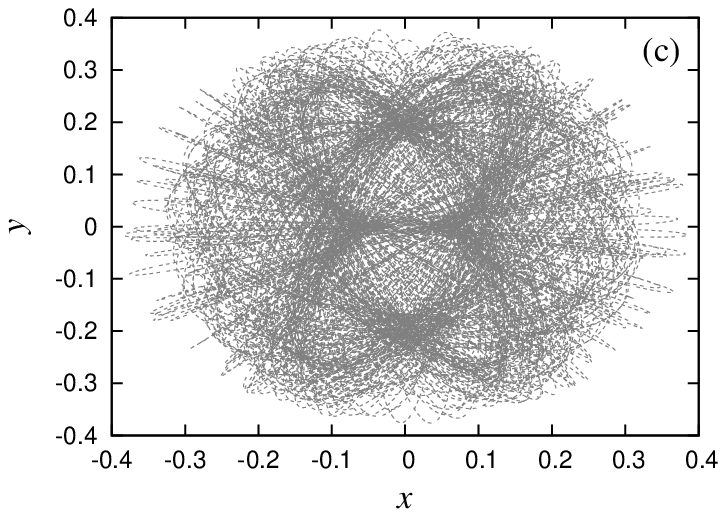}
\includegraphics[width=5.0cm]{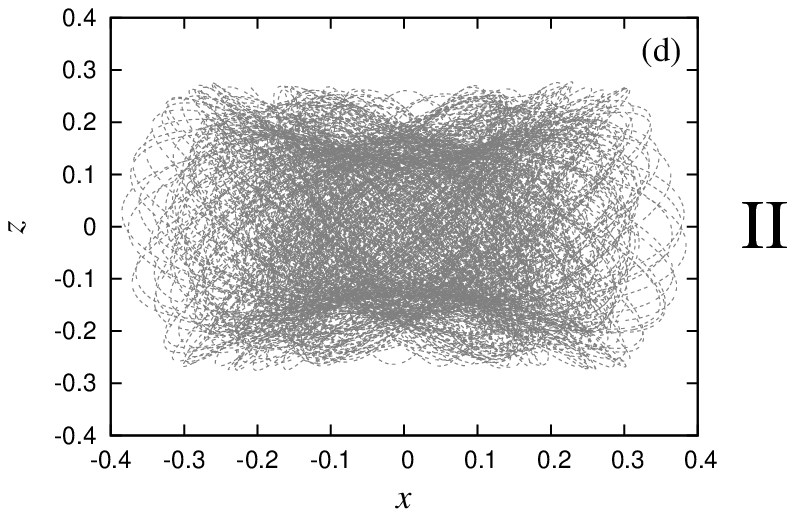}\\
\includegraphics[width=5.0cm]{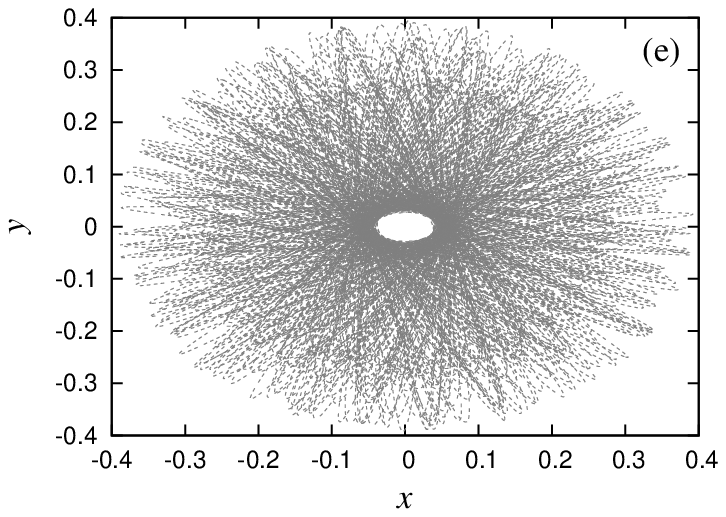}
\includegraphics[width=5.0cm]{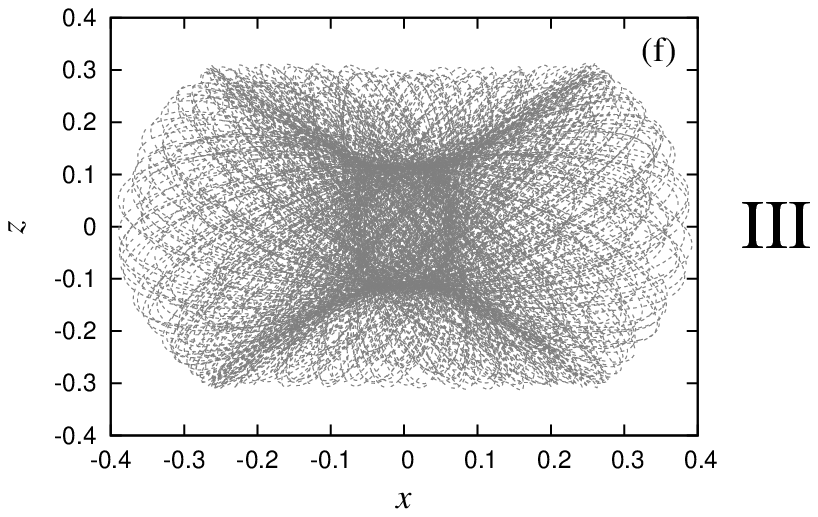}\\
\caption{Projections of orbit B on the $(x,y)$ (left column) and the
  $(x,z)$ plane (right column) for different time intervals: $0\leq t
  \le 2500$ (upper row, interval I), $10000< t \le 12500$ (middle
  row, interval II) and $17500< t \le 20000$ (lower row, interval
  III).}
\label{3DOForbitB}
\end{figure}

In figure \ref{3DOForbitB}
we show the projections on the $(x,y)$ (left column) and the $(x,z)$
plane (right column) of orbit B for three different time
intervals. Even by mere inspection one can observe the complexity of
the evolution of orbit B. Although it is not safe to make accurate
predictions for the nature of an orbit based on its form in
the configuration space, we can say that orbit B looks regular in
intervals I and III (upper and bottom rows of figure \ref{3DOForbitB}
respectively), while it appears more complicated in interval II
(middle row of figure \ref{3DOForbitB}). These observations suggest
that the orbit is initially regular and after a chaotic phase becomes
regular again.

\begin{figure}
\centering
\includegraphics[width=7.0cm]{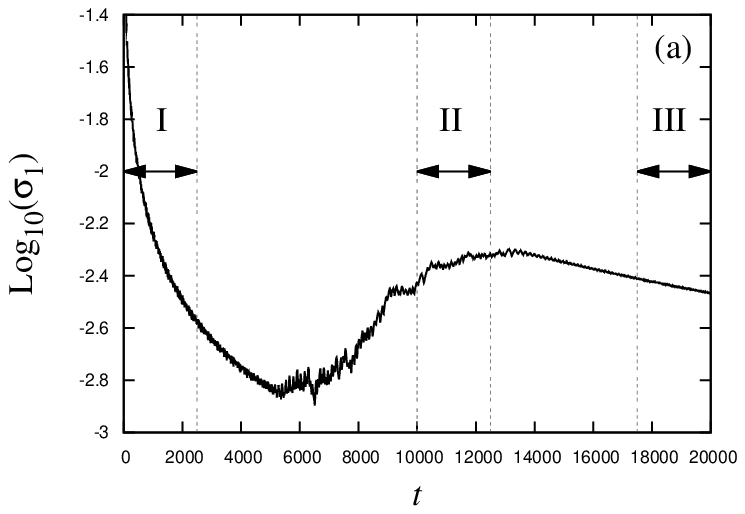}
\includegraphics[width=7.0cm]{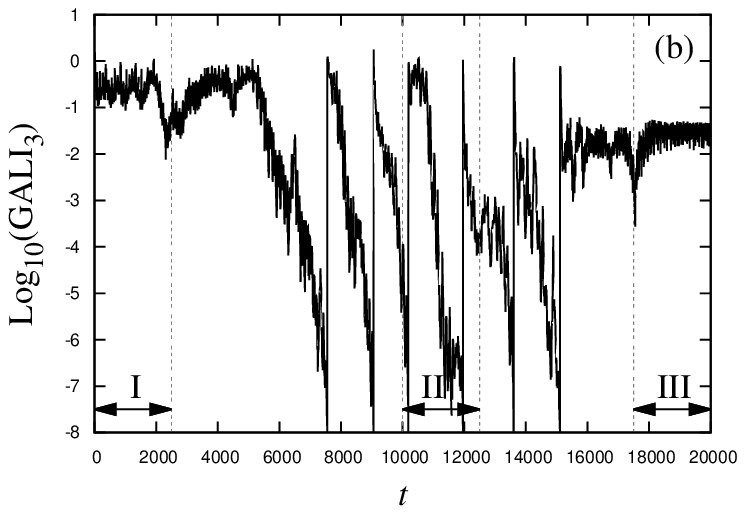}
\caption{Time evolution of the logarithm of (a) the finite time MLE
  $\sigma_1$, and (b) the reinitialized GALI$_3$ of orbit B. The three
  different time intervals I, II, and III, that correspond to the rows
  of figure \ref{3DOForbitB}, are located between the vertical dashed
  gray lines.}
\label{3DOF2_MLE_GALI}
\end{figure}

Exactly because orbit projections (especially for systems of more than
2 dof) are so difficult to interpret, we compute the $\sigma_1$
and the GALI$_3$ of orbit B, in order to analyze the
stages through which the orbit passes. In figure \ref{3DOF2_MLE_GALI}(a)
we see that $\sigma_1$ decays for $ t \lesssim 5000$, implying that
the orbit is regular, then increases to higher values, indicating
the possibility of chaotic motion, and finally for $t \gtrsim 14000$
decays again to zero, suggesting a return to regularity.

It is remarkable how clearly GALI$_3$ captures the different dynamical phases
of the orbit's evolution (recall that that whenever GALI$_3\leq 10^{-8}$, we set GALI$_3=1$ and repeat its computation using three orthonormal deviation vectors). GALI$_3$ initially oscillates around a high non-zero
value, asserting that the motion is regular until about $t\approx
5000$ when it starts to decrease rapidly to zero indicating that the
orbit is chaotic. This phase lasts until $t \approx 15000$, when GALI$_3$ jumps and remains practically constant until the end of the integration period, indicating that the orbit has become regular again.

By comparing the panels of figure \ref{3DOF2_MLE_GALI} we see that
$\sigma_1$ does not convincingly identify the transitions from regularity
to chaoticity and vice versa, for reasons that were already discussed.
Furthermore, we see that, even in the last interval of the
orbit's evolution ($t \gtrsim 14000$), where $\sigma_1$ begins to fall towards
zero, it decreases so slowly that the nature of the orbit is far from clear.
On the other hand, GALI$_3$ can successfully detect the local properties of the dynamics and provide us with a clear knowledge of the chaotic vs.~regular nature of an orbit, even for small time windows, where the orbit's nature often changes rapidly.

\begin{figure}
\centering
\includegraphics[width=10.0cm]{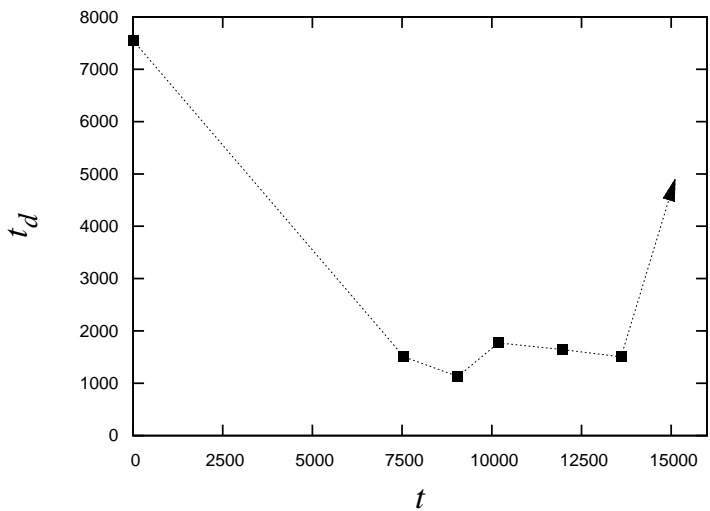}
\caption{The time $t_d$ that the reinitialized GALI$_3$ needs to
  decrease from GALI$_3=1$ to GALI$_3\leq 10^{-8}$, as a function of
  the integration time $t$ of orbit B. Note that for $t\gtrsim 15000$ GALI$_3$
  no longer falls to zero, indicating that the motion has entered a regular domain.}
\label{3DOF_GALI3zeroT}
\end{figure}

The distinction between the regular and chaotic intervals of orbit B is well depicted in figure \ref{3DOF_GALI3zeroT},
where the time $t_d$ needed for the reinitialized GALI$_3$ to become
$\leq 10^{-8}$ is plotted as a function of the integration time. Small
$t_d$ values for $7500 \lesssim t \lesssim 14000$ correspond to
chaotic epochs, where GALI$_3$ goes to zero exponentially fast,
while larger $t_d$ values correspond to intervals where GALI$_3$ takes longer to decay to zero. Observe also in figure \ref{3DOF_GALI3zeroT} the remarkable fact (shown by an upwardly pointing arrow) that, after $t\gtrsim 15000$, GALI$_3$ no longer falls to zero (see figure \ref{3DOF2_MLE_GALI}(b)) until the end of the integration time! This is certainly not expected in TI systems. It does occur, however, for orbit B of the 3 dof TD model, as well as
orbit C of a similar system (see below).

\subsubsection{A case where the bar gets stronger in time}\label{sec:mass_increase}

Let us now study the case of a linear increase of the bar's mass
$M_B$ from the initial value $M_B(t_0=0)=0.1$ to
$M_B(t_{final}=20000)=0.2$, as we did in the case of the 2 dof
model. We take again $\alpha=5 \times 10^{-6}$ in (\ref{MBlinlaw}). As
an example, we consider the orbit C with initial condition
$(x,y,z,p_x,p_y,p_z)=(0.225,0.0,0.25,0.0,0.0)$ and initial energy
$H_C(t_0)=E_C(t_0)=-0.441$, which undergoes an interesting sequence of
dynamical transitions. In this case, $Q_b$ starts from
$Q_b(t_0)=0.425$ and reaches the value $Q_b(t_{final})= 0.6732$ at the
end of the orbit's evolution.

\begin{figure}
\centering
\includegraphics[width=5.0cm]{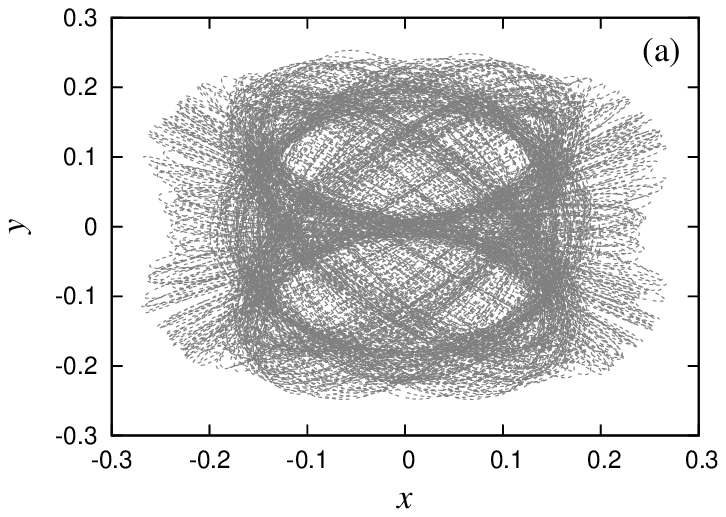}
\includegraphics[width=5.0cm]{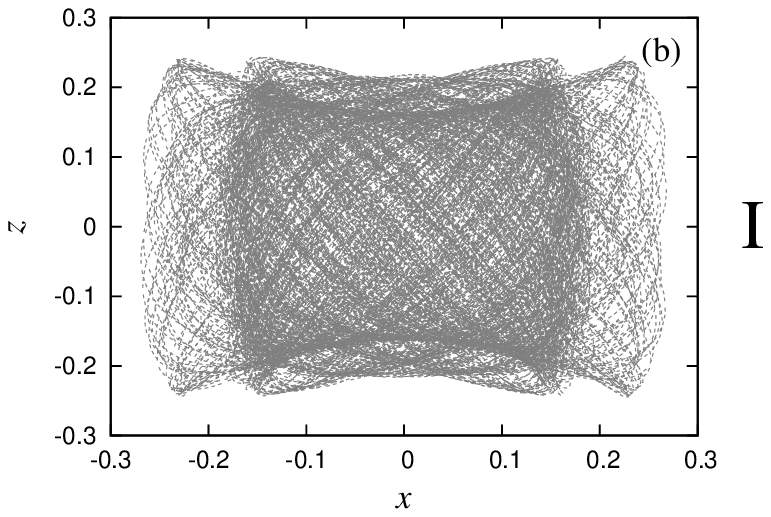}\\
\includegraphics[width=5.0cm]{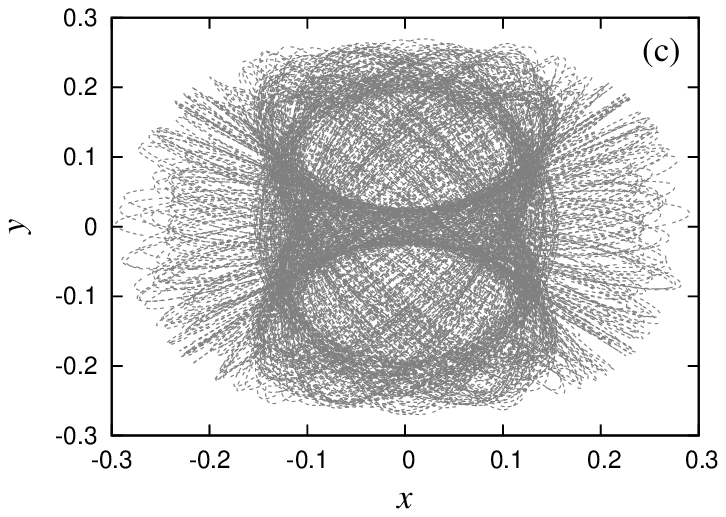}
\includegraphics[width=5.0cm]{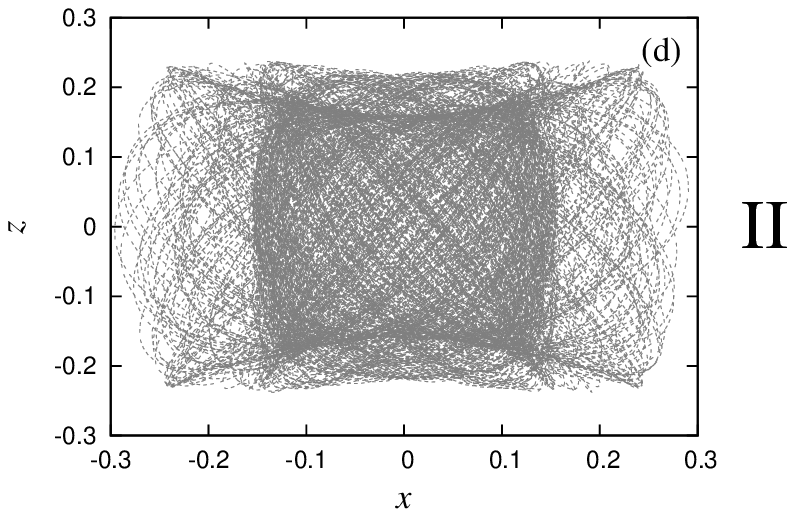}\\
\includegraphics[width=5.0cm]{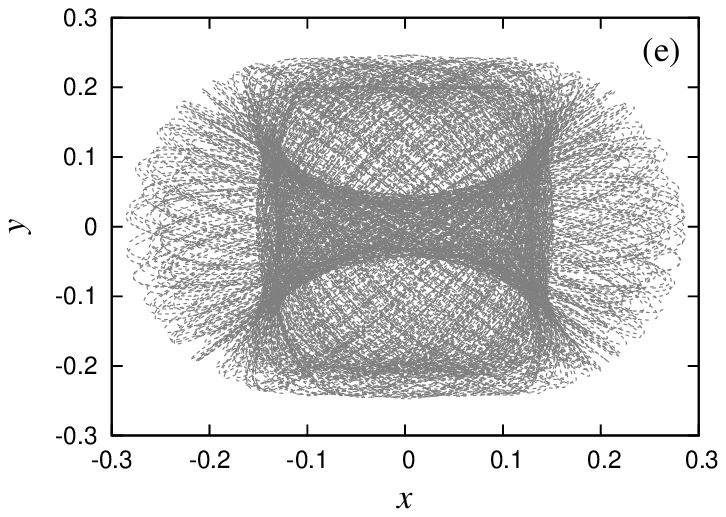}
\includegraphics[width=5.0cm]{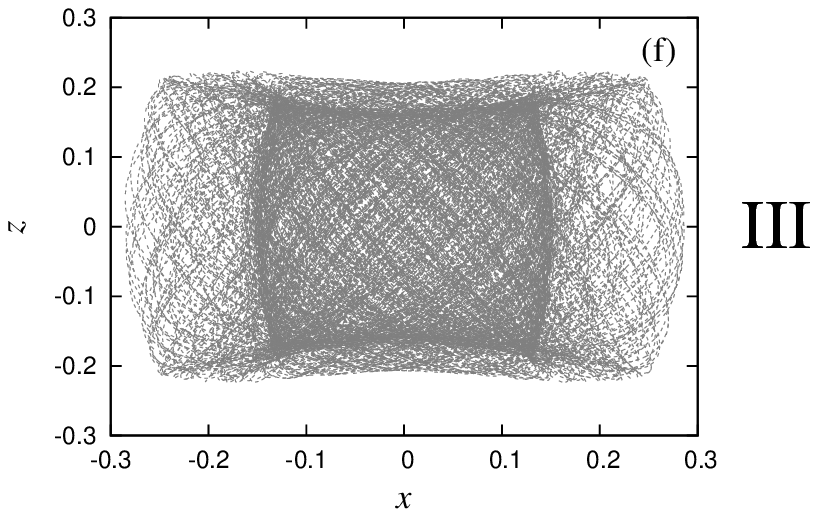}\\
\caption{Projections of orbit C on the $(x,y)$ (left column) and the
  $(x,z)$ plane (right column) for different time intervals: $0\leq t
  \le 2500$ (upper row, interval I), $5000< t \le 7500$ (middle
  row, interval II) and $17500< t \le 20000$ (lower row, interval
  III).}
\label{3DOForbitA}
\end{figure}

In figure \ref{3DOForbitA}
we plot the projections of orbit C on the $(x,y)$ and the $(x,z)$
planes (left and right column respectively) for three different time
intervals. The orbit looks more or less regular in intervals I and
III, although its shape is quite different in the two intervals. In interval
II it looks a bit more complicated and seems to represent a transition
between the two different configurations of intervals I and III. However, as
has already been mentioned, the mere inspection of the orbit is not
enough to accurately inform us about its chaotic or regular nature.

\begin{figure}
\centering
\includegraphics[width=7.0cm]{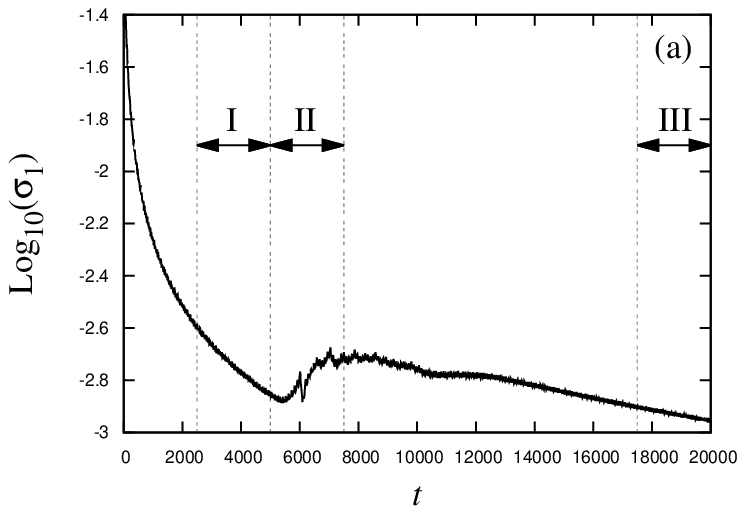}
\includegraphics[width=7.0cm]{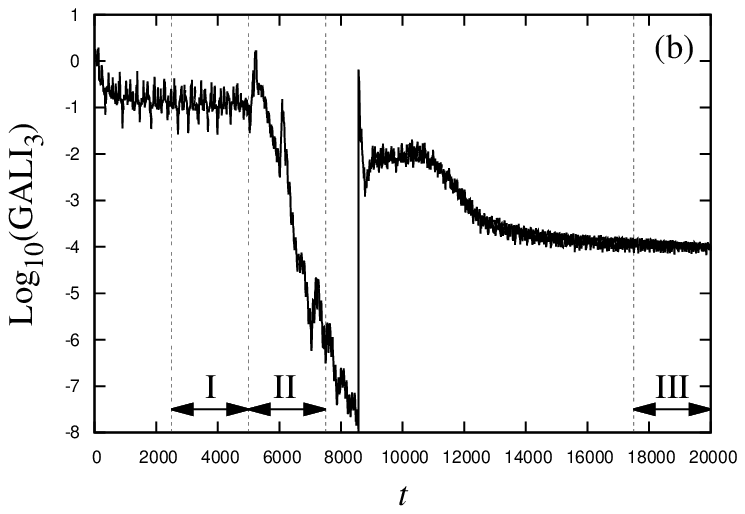}
\caption{Time evolution of the logarithm of (a) the finite time MLE
  $\sigma_1$, and (b) the reinitialized GALI$_3$ of orbit C. The three
  different time intervals I, II, and III, that correspond to the rows
  of figure \ref{3DOForbitA}, are located between the vertical dashed
  gray lines.}
\label{3DOF1_MLE_GALI}
\end{figure}
In figure \ref{3DOF1_MLE_GALI}(a) we plot the time evolution of
$\sigma_1$ for orbit C. From this figure we see that $\sigma_1$
initially decays to zero, suggesting the regular character of the
orbit. Then, at $t \approx 5500$ we observe a small increase of
$\sigma_1$, which indicates a dynamical change in the orbit's
behavior. This is soon followed by a monotonic decrease of
$\sigma_1$, which might indicate that orbit C becomes regular again. All this,
however, is highly speculative. By contrast, the
time evolution of the reinitialized GALI$_3$ (see figure
\ref{3DOF1_MLE_GALI}(b)) shows a lot more clearly the transition epoch
between the two different regular states. Initially GALI$_3$ remains
constant and different from zero, providing clear evidence that the
orbit is regular. Then, for $5000 \lesssim t \lesssim 8500$ it
decreases to $\leq 10^{-8}$ over a relatively long time interval,
indicating a fundamental change in the character of the orbit. Finally, after
reinitializing the index's computation at $t \approx 8500$, GALI$_3$ begins to converge to a positive constant, demonstrating the remarkable fact that the orbit has again become regular! Thus, here also, as in the case of orbit B (see figure \ref{3DOF2_MLE_GALI}(b)), an interlude of chaotic behavior is followed by a transition to regularity, which lasts until the end of our integration time.

\subsection{Global dynamics of the 3 dof model}\label{sec:global}

After establishing the efficiency of the GALI method in identifying time
intervals where an orbit of a TD model is regular or chaotic, let us
use it to study in a more global way the dynamics of Hamiltonian
(\ref{eq:Hamilton}). In particular, we will investigate the case
considered in section \ref{sec:mass_increase}, where the mass of the
bar component increases linearly from $M_B(t_0=0)=0.1$ to
$M_B(t_{final}=20000)=0.2$, corresponding to $\alpha=5 \times
10^{-6}$ in (\ref{MBlinlaw}).

In \cite{MA11} the TI version of model (\ref{eq:Hamilton})  was considered
for fixed values $M_B=0.1$ and $M_B=0.2$ respectively. In that work, these two
cases were referred as models `S' and `M' respectively. Ensembles of
50000 different initial conditions were integrated up to $t=10000$ time units,
and the GALI method was used to accurately determine the percentages of chaotic orbits. The analysis performed in \cite{MA11} showed that chaotic behavior is more dominant for the `M' model (i.e.~the one with the more massive bar component). Our TD model (\ref{eq:Hamilton}) coincides with model `S' of \cite{MA11} at $t=0$ and becomes model `M' at $t=20000$. Thus, it is of interest to check for this model if the same sets of initial conditions considered in \cite{MA11} show a tendency to increase their chaoticity as time grows from $t=0$ to $t=20000$, in agreement with the general trend
found in \cite{MA11}. For this reason, we evolve the same three classes of initial condition distributions considered in \cite{MA11}:
\begin{itemize}
\item distribution $I$:\ 5000 orbits equally spaced in the space
  $(x,z,p_{y})$ with $x\in [0.0,7.0]$, $z\in [0.0,1.5]$, $p_y \in
  [0.0,0.45]$ and $(y,p_{x},p_{z})=(0,0,0)$,\
\item distribution $II$:\ 5000 orbits equally spaced in the space
  $(x,p_{y},p_{z})$ with $x\in [0.0,7.0]$, $p_y\in [0.0,0.35]$, $p_z
  \in [0.0,0.35]$ and $(y,z,p_{x})=(0,0,0)$,\
\item distribution $III$:\ 5000 orbits whose spatial coordinates are
  chosen randomly over the mass density distribution of model `S'
  (according to the so-called rejection method) within the rectangular
  box $-a\leq x \leq a$, $-b\leq y \leq b$, $-c\leq z \leq c$, with
  $(p_{y},p_{z})=(0,0)$, and $p_x>0$ obtained from (\ref{eq:Hamilton})
  with $H$ taking for each initial condition a fixed random value in
  the interval $[-0.22,0]$ (see \cite{MA11} for more details).
\end{itemize}
The only difference between the distributions of our study and those of
\cite{MA11} is that we use only 5000 initial conditions, instead of
50000 used in \cite{MA11}, in order to facilitate our computations.

\begin{figure}
\centering
\includegraphics[width=10.0cm]{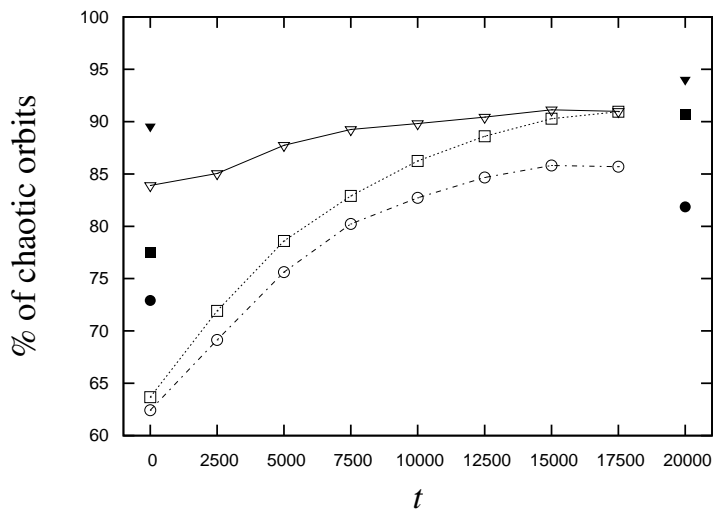}
\caption{Time evolution of the percentages of chaotic orbits for the
  initial distributions $I$ (dotted line), $II$ (dashed line), $III$
  (solid line) of model (\ref{eq:Hamilton}), when $M_B$ is linearly
  increased from $M_B(t_0=0)=0.1$ to $M_B(t_{final}=20000)=0.2$ (see
  text for more details). The percentages are obtained for successive
  time windows of length $\Delta t=2500$ time units, and are
  attributed to the starting times of these intervals. For comparison,
  the percentages of the TI models `S' (with $M_B=0.1$) and `M' (with
  $M_B=0.2$) obtained in \cite{MA11}, are also plotted at $t=0$ and
  $t=20000$ respectively, by filled squares (distribution $I$),
  filled circles (distribution $II$), and filled triangles
  (distribution $III$).}
\label{3DOF_GALIpercICs}
\end{figure}

Since our model is TD the percentage of chaotic orbits of the three
orbital distributions is expected to change in time. In order to
monitor these changes we adopt the following strategy: We divide the
total integration time of 20000 time units in eight successive time
windows of length $\Delta t=2500$ time units. At the beginning of
each time window, we reinitialize GALI$_3$ to unity and follow
a new set of three orthonormal deviation vectors for each orbit. Then for
each time window we calculate the ``current percentage of chaotic
orbits'' as the fraction of orbits whose GALI$_3$ becomes $\leq
10^{-8}$ in the duration of the time window. The results of this
procedure are shown in figure \ref{3DOF_GALIpercICs},
where the percentages obtained for each time window are attributed to
its starting time. The percentages found in \cite{MA11} for models `S'
and `M' are also plotted at $t=0$ and $t=20000$ respectively, by
filled squares (distribution $I$), filled circles (distribution $II$),
and filled triangles (distribution $III$).

In accordance with the results of \cite{MA11} we see an increase of the
fraction of chaotic orbits for all three ensembles of
orbits. Distribution $III$ contains more chaotic orbits among the
three ensembles, while the percentages of chaotic orbits is smallest
for distribution $II$, in agreement with what was observed in
\cite{MA11}. Obviously, one should not expect to find the same
numerical values with the percentages reported in \cite{MA11}. For
example, our TD model is identical to model `S' of \cite{MA11} only
momentarily for $t=0$, and thus the percentages of the TD model
reported for $t=0$ do not correspond exactly to model `S'. Apart from
the fact that the percentages reported in \cite{MA11} were obtained
for a TI model, another difference is that in \cite{MA11} each orbit was
integrated in a fixed Hamiltonian system for 10000 time units, while in our
study each orbit is integrated for only 2500 time units in a TD potential.
Nevertheless, the dynamical trends obtained in our study are in good agreement with the ones presented in \cite{MA11}, clearly showing the efficiency of
the GALI method in analyzing TD models.

\section{Conclusions}\label{sec:conclusions}

Autonomous Hamiltonian systems are conveniently studied for  fixed values of the total energy, where the location and extent of their regular and chaotic regions are time independent and can be accurately identified by a variety of methods especially in the low degree of freedom case. Even in such TI systems the dynamics can exhibit remarkable complexity, as there exist regimes of ``strong'' and ``weak'' chaos, as well as varying degrees of regularity, as the motion can occur on invariant tori of different dimensions and exhibit surprising localization properties in configuration and frequency space \cite{BS12}.

Naturally, therefore, Hamiltonian systems which are explicitly TD are expected to be a lot more complicated, since, in the absence of TI integrals, all the above attributes evolve in time. For example, in the TI case, orbits do not change their nature: If they are initially regular they will always remain so, while if chaotic they can get trapped for long times on the boundary of regular regimes (exhibiting a ``weak'' form of chaos), but will never entirely relinquish their chaoticity. This is not so in the TD case, where individual trajectories may indeed display sudden transitions from regular to chaotic behavior and vice versa during their time evolution.

In the present paper, we have sought to shed some new light on these fascinating phenomena by studying the dynamics of a mean field model of a barred galaxy, whose mass parameters are allowed to vary linearly in time. Our primary goal was to show that transitions from order to chaos and vice versa do occur in this model and can be monitored much more accurately by local methods such as the GALI spectrum, rather than the more traditional approach of LEs. In addition, we wanted to investigate some astronomical properties of this TD system as it does incorporate some of the features appearing in N-body simulations and TI analytic potentials.

In this regard, we have chosen for simplicity to vary only two parameters in time, $M_B$ and $M_D$, keeping the size of the bar and the pattern speed fixed. Since the total mass is constant, whatever mass the bar loses is gained by the disc component and vice-versa. Moreover, to investigate more thoroughly the observed dynamical transitions, we have extended the maximal integration time of the orbits to $T=20000\;Myr$ (20 billion years), which corresponds to a total interval of nearly 2 Hubble times. However, our results demonstrate that rich behavior and fundamental changes can also be observed within a single Hubble time of $10000\;Myr$.

Most importantly, the GALI method turns to be again very efficient and accurate in the detection of chaotic motion in a TD system, as in the case of TI models. Furthermore, our work reveals that the method is especially suited for detecting intervals where an orbit changes its state fundamentally. By following the times that the GALIs require to fall to zero, one can describe in detail the orbit's successive passages from order to chaos and vice versa.

Finally, we focus on a more global astronomical study, where one would like to estimate qualitatively and quantitatively the relative fraction of regular and chaotic motion in such galaxy models. To this end, we choose different sets of initial conditions, launch them in phase space and classify regular and chaotic orbits, depending on whether the GALI fluctuates around a non-zero value or falls exponentially to zero. We have thus been able to verify that the conclusion of an earlier publication on the TI model \cite{MA11}, that the percentage of chaos grows as the mass of the bar increases, holds true in the TD case as well. It would be highly interesting to investigate these questions in more realistic models, where besides the mass parameters the rotation frequency of the galaxy is also allowed to vary accordingly.

In closing, it is important to point out the advantages of the GALI method over the computation of the finite time MLE, as the GALIs do succeed in clearly capturing the transitions between different regular states and identifying the intermediate chaotic phases. By contrast, the manifestation of these different dynamical behaviors is much less pronounced in the time evolution of the MLE described by $\sigma_1$ in this paper. Evidently, the practice of averaging Lyapunov exponents over an orbit's history smoothens out their fluctuations over short-lived events and gives them meaning only in the sense of the long time limit.

\section*{Acknowledgments}

This work was partially supported by a grant from the GSRT,
Greek Ministry of Development, for the project ``Complex Matter'',
awarded under the auspices of the ERA Complexity Network. It is also supported by the European Union (European Social Fund) and Greek national funds through
the Operational Program ``Education and Lifelong Learning'' of the
National Strategic Reference Framework (NSRF) - Research Funding
Program: THALES. Investing in knowledge society through the European
Social Fund. T.~M.~would like to thank Dr.~I.~Martinez-Valpuesta and
Dr.~R.~Machado for their fruitful comments and discussions on this
work. Ch.~S.~is grateful for the hospitality of the Max Planck
Institute for the Physics of Complex Systems in Dresden, Germany,
during his visit in July - August 2012, when part of this work was
carried out.


\end{document}